# HABITABILITY OF POLAR REGIONS IN TIDALLY LOCKED EXTRASOLAR PLANET NEAR THE M-DWARF STARS

*Nishith Burman*

Northeastern University

## ABSTRACT

Since the launch of Kepler and Hubble more than a decade ago, we have come a long way in the quest to find a potentially habitable exoplanet. To date, we have already discovered more than 4000 exoplanets most of which are not suitable for sustaining life. Of all those that can potentially sustain life, A large number has been found rotating synchronously around their parent star, mostly Red dwarf star. Due to their synchronous rotation, these planets receive very uneven stellar heating. Synchronous rotation of these planets causes one side of the planet to permanently face the parent star while the other side remains dark. This results in an extreme climatic condition that is not feasible for sustaining life. Although these theories about exoplanets are well known, a systemic study of habitability of polar regions within an exoplanet using different climate models has not been done yet. Here I review the current literature on tidal locking and its impact on habitability and introduce the concept of habitability in the poles of these exoplanets. I focus on my understanding of the climatic condition in the polar region of the earth and based on that I present the concept of habitability in the poles of these exoplanets.

## 1. INTRODUCTION

Habitability of exoplanets depends on different factor ranging from distance from the host star, Mass, Radii (wright et al. 2018), eccentricities (Wang et al. 2017), orbital dynamic (Deitrick et al. 2017), synchronous rotation (Barnes Rory, 2008), continent carbon-silicate weathering (Lewis et al. 2018), magnetic fields (Turner et al. 2018), tidal heating (Wu and Goldreich, 2002; Mardling and Lin, 2002; Rodriguez et al. 2012; Van Laerhoven et al. 2014) and Ocean heat transport (Hu et al. 2014), etc. Of all these factors, one of the most common factors that affect almost all the potentially habitable exoplanets is synchronous rotation. Synchronous rotation leads to tidal locking (Rory Barnes, 2017). Tidal locking is a phenomenon that results in an extreme climatic condition in almost every region of the planets.

Figure 1. The below figure shows the images of the distribution of potentially habitable exoplanets around different kinds of star systems. Most of these planets are tidally locked with their host stars.

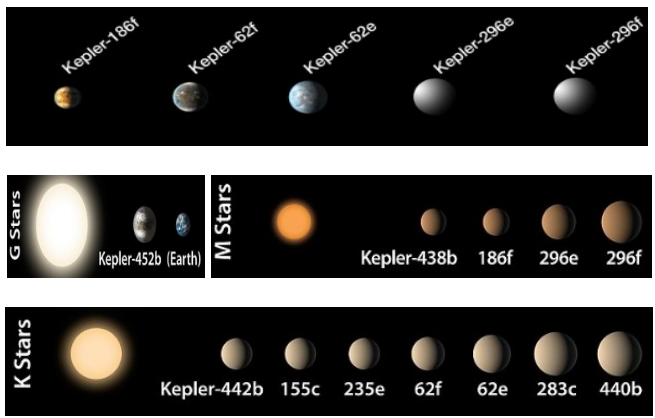

Tidal locking (Rory Barnes, 2017) leads to strong heating of a planet at a substellar region which can change or even control weathering on the planet. The change in weathering, in turn, can destabilize the climatic condition within the planet. These drastic climate effects could make planets that otherwise have the potential for life to instead be uninhabitable. The substellar regions being consistently close to the star would receive more sunlight and heat. This causes increased weathering in the substellar region and this leads to the rapid erosion of the atmosphere. This entire process of enhanced weathering and enhanced erosion of atmosphere in the substellar region is commonly referred to as enhanced substellar weathering instability (ESWI) (Nola Taylor, 2011; Kite et al. 2011; Nikku Madhusudan). Similarly, if there is a cooling of the substellar region due to any reason, the weathering process may slow down and gases may build up. Volcanic activities can further eject more material into the atmosphere which can lead to a runaway greenhouse effect (Ramirez et al., 2017) and thus this can also contribute to the erosion of atmosphere.

With the development of 3-dimensional global climate models (GCMs), the surface properties of synchronously rotating planets can be explored more self-consistently. The first models were relatively simple but found that synchronously rotating planets can support liquid water (Joshi et al. 1997). More recent investigations have confirmed this result (Yang et al. 2013, Wordsworth et al. 2011, Way et al. 2015, Pierrehumbert 2011, Kopparapu et al. 2016, Shield et al. 2016).

So, if the substellar region is below the ocean surface or if there is not enough surface exposed to weathering or if the gases that are absorbed during weathering is not prevalent in the atmosphere, In all of the above cases, the atmosphere will not corrode and for all such cases even if the temperature is inhospitable for most of the regions, the polar region will experience stellar radiation that is not as intense as the substellar region and not as mild as the dark region. In the next section, I explore the notion of habitability in the polar regions of these planets in detail.

## 2. PROPOSED WORK

### 2.1. PREVIOUS WORK

A lot of work has been carried out previously to find out the impact of different physical and chemical conditions on the habitability of these exoplanets. For instance, Wang et al. (2017) used a 3-Dimensional atmospheric general circulation model to study the effect of eccentricities on climate and habitability of M-dwarf exoplanets. They found out the eccentricities could drastically change the climate pattern which could make the planet inhabitable. Similarly, Yongyun Hu et al. (2014) used a coupled atmosphere-ocean model to study the role of ocean heat transport on the climates of tidally locked M-dwarf exoplanets. They found out that ocean heat transport can substantially extend the dayside habitable area and efficiently warm the nightside so that atmospheric collapse does not occur. They also found out If the greenhouse effect and stellar radiation are strong enough, ocean heat transport can even cause global deglaciation. Most of the above model discusses the impact of the different climatic conditions on the habitability of the substellar region and dark region. But none of them discusses the habitability of the polar region.

### 2.2. My APPROACH

To understand the habitability of polar regions in exoplanets, we must understand the habitability of poles in Earth's context.

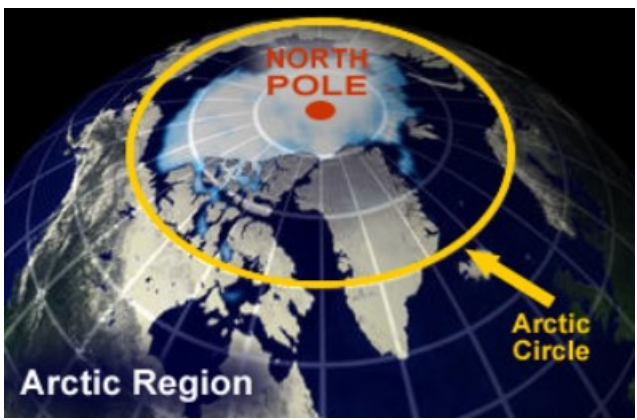

Figure 2. above shows the visualization of the Arctic region which is located in the North pole.

Both Arctic and Antarctic region is cold throughout the year because they receive a very small amount of solar radiation, due to the low average incidence of the direct solar radiation at high altitudes. Solar radiation received by the polar region is only about 30% of the radiation received by the equator, during the winter radiation received by the poles even falls to zero. In fact, the annual total solar radiation received by poles is about 15% of that received by the equator. Based on the above data we can infer that even if dayside of these planets are extremely hot the poles would receive far less radiation and thus the temperature of the poles will be far less hostile. Similarly, In comparison to the nightside, it will receive a far greater amount of radiation and thus again the climate and environment of these regions would be far more habitable.

Figure 3. below shows the incidence of direct radiation on the different regions of the planet.

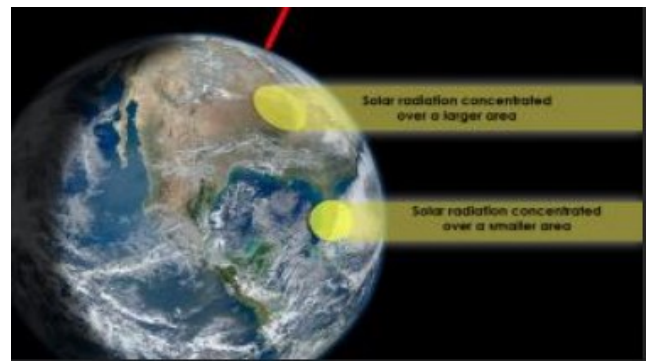

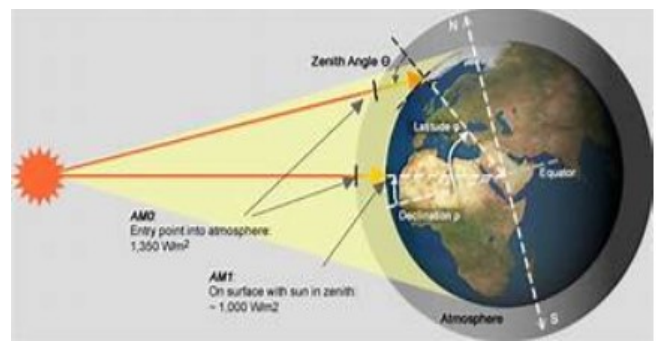

To understand the effectiveness of this theory let's take the example of exoplanet Kepler – 438b orbiting around the M-type Red Dwarf star Kepler-438. The mass and radii of Kepler-438 are 0.54 and 0.52 time respectively the mass of Sun. The surface temperature of this star is nearly 3748 K. The mass of this planet is 1.46 times the mass of the Earth and the orbital period of this planet is 35.2 days. Kepler-438b has a rocky surface and is within the habitable zone of this star. But the age of the Red dwarf star is 4.8Gyr and at this point, it is highly active and it constantly buffets the planet with a huge amount of radiation and right now this makes the Kepler-438b Inhospitable. But eventually, with age, the star's activity will decrease and this would increase the probability of Kepler-438b being more habitable. So, this makes it a candidate worth researching. Now, based on the current technology we have, It is expected that the surface temperature of the planet is around 140 F. Although 140 F is quite an extreme temperature, the primordial soup in which the first amino acid and Nucleotides which are the basic building block of life were generated had a similar temperature. Also, 140 F is the average temperature of the planet which means the surface temperature will be different for the different region and there is a high probability that the surface temperature of the polar region will be considerably less than 140 F as far less amount of radiation will be incident in this region of the planet and thus there will be a very high probability that this region would be the most suitable region for sustaining the life.

## 5. CONCLUSION

Tidally Locked Exoplanets rotating around a stable red dwarf star and having conditions similar to the exoplanet Kepler – 438b are good candidates for planets suitable for sustaining life. These planets are rocky, the temperature is moderate at the polar regions but harsh in the substellar and the dark regions and it is highly likely that they can support water. Although these planets may be buffeted by the solar flares coming out of their host stars during the first 8 Gyr – 10 Gyr, but once this phase passes away, these stars become far more stable and thus the probability that poles of these planets can sustain life increases considerably. Of Course, based on the different geological conditions one of the poles can have a more extreme climate and environment. For instance, In the case of Earth, Antarctica has far more severe climatic conditions than Artic due to its geological structure and thus only a handful of organisms are capable of surviving those harsh climatic conditions. However, regardless of all the other aspects, the polar region of tidally locked exoplanets present at the inner edge of the Goldilocks zone receives consistently low radiation throughout the year and thus may have the temperature that may be just suitable enough to harbor life.